\documentclass[aps,prd,preprintnumbers,showpacs,showkeys,nofootinbib,
superscriptaddress,fleqn,floatfix,tightenlines, preprint]{revtex4-1}
\usepackage{amsmath,amsfonts,amssymb,amscd,amsxtra,amsthm}
\usepackage{graphicx,slashed,bm}
\usepackage{mathrsfs,multirow,booktabs,epstopdf}
\usepackage[utf8]{inputenc}
\usepackage[normalem]{ulem} 
\usepackage[dvipsnames]{xcolor} 
\renewcommand\sout{\bgroup \color{red} \ULdepth=-.5ex \ULset}

\begin{document}
\preprint{INHA-NTG-06/2016}
\title{$K^0 \Lambda$ and $D^- \Lambda_c^+$ production induced by pion
  beams off the nucleon}
\author{Sang-Ho Kim}
\email[E-mail: ]{sangho.kim@apctp.org}
\affiliation{Asia Pacific Center for Theoretical Physics (APCTP), Pohang,
Gyeongbuk, 790-784, Republic of Korea}
\author{Hyun-Chul Kim}
\email[E-mail: ]{hchkim@inha.ac.kr}
\affiliation{Department of Physics, Inha University, Incheon 402-751,
Republic of Korea}
\affiliation{School of Physics, Korea Institute for Advanced Study (KIAS),
Seoul 130-722, Republic of Korea}
\author{Atsushi Hosaka}
\email[E-mail: ]{hosaka@rcnp.osaka-u.ac.jp}
\affiliation{Research Center for Nuclear Physics (RCNP), Osaka 
University, Ibaraki, Osaka, 567-0047, Japan}
\affiliation{J-PARC Branch, KEK Theory Center, Institute of Particle and 
Nuclear Studies, KEK, Tokai, Ibaraki, 319-1106, Japan}
\date{\today}
\begin{abstract}
We present a comparative study of the pion induced production of
$K^0 \Lambda$ and $D^- \Lambda_c^+$ off the nucleon.
A hybrid framework is utilized by combining an effective Lagrangian
method with a Regge approach. We consider the $t$-channel process in a
plannar diagram by vector-meson Reggeon exchanges and the $u$-channel
one in a non-planar diagram by baryon Reggeon exchanges.
The present model reproduces the $K^0 \Lambda$ production data well
with a few parameters. Having fixed them, we predict the $D^- \Lambda_c^+$
production, which turns out to be about $10^4-10^6$ times smaller than
the strangeness one, depending on the kinematical regions.
\end{abstract}
\maketitle
\section{Introduction}
Experimental findings of new heavy hadrons have renewed great interest
in heavy-quark physics~(see, for example, following 
reviews\cite{Swanson:2006st, Voloshin:2007dx, Brambilla:2010cs,
  Chen:2016qju}). For example, the Belle Collaboration, BABAR
Collaboration, BESIII Collaboration, and LHCb Collaboration have
reported new types of heavy mesons~\cite{Choi:2003ue, Aubert:2004ns,
  Aubert:2005rm, Abe:2007jna, Choi:2007wga, Belle:2011aa,
  Ablikim:2013mio, Liu:2013dau, Ablikim:2013wzq, Aaij:2013zoa,
  Aaij:2014jqa}. Moreover, a new bottom baryon $\Xi_b^0(5945)$ has
been observed by the CMS Collaboration~\cite{Chatrchyan:2012ni}, and
two new bottom baryon resonances $\Xi_b'(5935)$ and $\Xi_b^*(5955)$
have been announced by the LHCb
Collaboration~\cite{Aaij:2014yka}. Recently, a 
new proposal was submitted to Japan Proton Accelerator Research
Complex (J-PARC) to measure the production of charmed hadrons.
The high-momentum pion beam line up to 20 GeV will be constructed to
produce excited charmed
baryons~\cite{Noumi:2014vfa,Shirotori:2014nua}. The $\pi^- p \to
D^{*-} \Lambda_c^+$ reaction was suggested as the first experiment and
the relevant theoretical works have been discussed recently by us in
collaboration with other authors~\cite{Kim:2014qha,Kim:2015ita}. 
Reference~\cite{Kim:2014qha} estimated the production rates of charmed 
baryons $Y_c^+$ in the process $\pi^- p \to D^{*-} Y_c^+$, where $Y_c^+$ is 
the $\Lambda_c^+, \Sigma_c^+$ and their excited states, with the help of 
quark-diquark model. Reference~\cite{Kim:2015ita} has predicted the
magnitude of the total cross section of the $\pi^- p \to D^{*-}
\Lambda_c^+$ in comparison with the process of its strangeness partner  
$\pi^- p \to K^{*0} \Lambda$ based on an effective Lagrangian method
and a hybridized Regge model. 

In the present work, we extend the previous
investigation~\cite{Kim:2015ita} to the reaction $\pi^- p \to D^-
\Lambda_c^+$ together with the $\pi^- p \to K^0\Lambda$ using a
similar method. We first concentrate on the $\pi^- p \to K^0\Lambda$ 
where experimental data exist, so that we are able to fix the
parameters such as coupling constants and scale factors.
Since the amplitudes have exactly the same strtucture, the
corresponding charm production $\pi^- p \to D^- \Lambda_c^+$ can be
easily studied, the parameters being assumed to be the same as the
strangeness case. We also examine the sensitivity of the results to
the changes of the parameters. To fix the Regge parameters such as
Regge trajectories and energy scale parameters, we use the
non-perturbative Quark-Gluon String Model (QGSM) 
developed by Kaidalov et al.~\cite{Kaidalov:1980bq,Boreskov:1983bu,
Kaidalov:1986zs,Kaidalov:1994mda} as was done in
Ref.~\cite{Kim:2015ita}. We find that the $D\Lambda_c^+$ ($K\Lambda$)
production is governed by vector-meson $D^*$ ($K^*$) Reggeon exchange
at forward angles, whereas by baryon $\Sigma_c$ ($\Sigma$) Reggeon
exchange at backward angles. We compare the production cross sections
for the $\pi^- p \to D^- \Lambda_c^+$ with other
models~\cite{Barger:1975fua,Kofler:2014yka}. 

The present work is organized as follows: In the next Section, we
explain the general formalism of the hybridized Regge model and fix
model parameters such as the coupling constants, Regge trajectories
and energy scale parameters, and the scale factors. In Section III, we 
show the numerical results and compare them with those from other
models. The last Section is devoted to the summary and conclusion. 

\section{General Formalism}
\begin{figure}[h]
\centering
\includegraphics[scale=0.53]{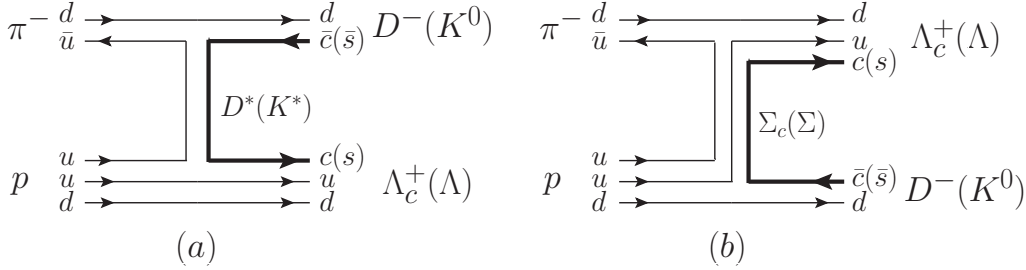}
\caption{(a) Planar and (b) nonplanar diagrams for the 
$\pi^- p \to D^- \Lambda_c^+\,(K^0 \Lambda)$ reaction.}
\label{FIG1}
\end{figure}

We start with explaining a hybridized Regge model combining with an
effective Lagrangian method. In general, there are two different
diagrams for both the reactions $\pi^- p \to D^- \Lambda_c^+$ and
$\pi^- p \to K^0 \Lambda$ as drawn in Fig.~\ref{FIG1}. Vector Reggeon
exchange can be understood as a planar diagram, so that the 
corresponding Regge parameters can be determined explicitly by
employing the QGSM~\cite{Kaidalov:1980bq,
  Boreskov:1983bu, Kaidalov:1986zs, Kaidalov:1994mda}.
On the other hand, the $u$-channel diagram for $\Sigma_c(\Sigma)$
Reggeon exchange is a nonplanar diagram for which there is no
theoretical ground on fixing the parameters. Thus, we will utilize
relevant phenomenologies to determine them.  
The incoming momenta of the pion and the proton are designated 
respectively by $k_1$ and $p_1$, and the outcoming momenta of the 
pseudoscalar meson and the $\Lambda_c(\Lambda)$ 
by $k_2$ and $p_2$, respectively.
We first investigate the strangeness production $\pi^- p \to K^0 \Lambda$ 
with the Regge parameters fixed. Then we will extend the same method to 
the charm production $\pi^- p \to D^- \Lambda_c^+$.

\subsection{Strangeness production $\pi^- p \to K^0\Lambda$: 
$K^*$ Reggeon exchange}
The $K^*$-exchange amplitude for the $\pi^- p \to K^0 \Lambda$
reaction is derived, based on the effective Lagrangians given as 
\begin{align}
\mathcal L_{\pi K K^*} &= 
-ig_{\pi K K^*} ( \bar K \partial^\mu \bm{\tau}\cdot\bm{\pi} K^*_\mu - 
               \bar K_\mu^* \partial^\mu \bm{\tau}\cdot\bm{\pi} K ),
                         \cr 
\mathcal L_{K^* N \Lambda} &=
-g_{K^* N \Lambda} \bar N \left[ \gamma_\mu \Lambda - 
\frac{\kappa_{K^* N \Lambda}}{M_N+M_\Lambda} \sigma_{\mu\nu} \Lambda 
\partial^\nu \right] K^{*\mu} + \mathrm{H.c.},
\label{eq:Lag1}
\end{align}
where $\pi,\,K$, and $K^*$ denote the fields corresponding to
the $\pi(140,0^-)$, $K(494,0^-)$, and $K^*(892,1^-)$ mesons
respectively, while $N$ and $\Lambda$ stand for the nucleon and
$\Lambda(1116,1/2^+)$ hyperon, respectively. The coupling constant
$g_{\pi K K^*}$ is calculated by using the experimental data on the
decay width $\Gamma(K^* \to K \pi)$~\cite{Olive:2014zz}: $g_{\pi K
  K^*} = 6.56$, whereas the $K^* N \Lambda$ coupling can be taken from
the Nijmegen soft-core potential (NSC97a)~\cite{Stoks:1999bz}  
\begin{align}
g_{K^* N \Lambda} = -4.26, \;\;\;\; \kappa_{K^* N \Lambda} = 2.91.
\label{eq:Coupl1}
\end{align}

To derive the $t$-channel Regge amplitude, we employ a hybridized
approach by replacing the Feynman propagator for vector-meson exchange
in the $t$ channel with the Regge propagator arising from the
corresponding Regge trajectory~\cite{Kim:2015ita,Donachie2002}   
\begin{align}
T_{K^*} (s,t)& = C_{K^*}(t) \mathcal M_{K^*}(s,t) 
\left( \frac{s}{s_{K^*}^{\pi N : K \Lambda}} \right)^{\alpha_{K^*}(t)-1} 
\Gamma (1-\alpha_{K^*}(t)) \alpha_{K^*}'.
\label{eq:RegAmpl1}
\end{align}
In doing so, we need to examine the behavior of the amplitude in the 
high-energy region, which will be discussed later. 
The amplitude $\mathcal M_{K^*}$ in Eq.~(\ref{eq:RegAmpl1}) without
the propagator is obtained from the Lagrangians in Eq.~(\ref{eq:Lag1})   
\begin{align}
\mathcal M_{K^*}(s,t) = g_{\pi K K^*} g_{K^* N \Lambda} \bar u_\Lambda 
k_1^\mu \left[ -g_{\mu\nu} + \frac{q_{t\mu} q_{t\nu}}{M_{K^*}^2} \right]
\left[ \gamma^\nu - \frac{i\kappa_{K^* N \Lambda}}{M_N+M_{\Lambda}} 
\sigma^{\nu \lambda }q_{t\lambda} \right] u_N,
\label{eq:Ampl1}
\end{align}
where $u_N$ and $u_\Lambda $ stand for the Dirac spinors of the initial 
nucleon and the final $\Lambda$, respectively.
The momentum transfer in the $t$ channel is defined as $q_t =
k_2-k_1$. Note that the $q_t^\mu q_t^\nu$ term in Eq.~(\ref{eq:Ampl1})
provides a certain contribution, which is distinguished from
$K^*$ exchange in $K^+\Lambda$ photoproduction $\gamma p \to K^+
\Lambda$~\cite{Guidal:1997hy} and $\pi^- p \to K^{*0}
\Lambda$~\cite{Kim:2015ita} reaction in which there is no contribution
from it. This is due to the fact that the antisymmetric tensor 
$\epsilon^{\mu\rho\alpha\beta}q_{t\rho}$ involved in the relevant Lagrangians
for $K^*$ exchange eliminates the $q_{t\mu} q_{t\nu}$ term by
contraction. The $C_{K^*}(t)$ represents a scale
factor~\cite{Kim:2015ita}, which is defined as 
\begin{align}
C_{K^*}(t) = \frac{a_t}{(1 - t / \Lambda^2)^2}.
\label{eq:scalFac_t}
\end{align}
It plays the role of a form factor that reflects a finite size of
hadrons. We choose $\Lambda=1\, \mathrm{GeV}$ and the parameter $a_t$
is determined by the experimental data at high energies: $a_t=0.40$. 

The differential cross section $d\sigma/dt$ is expressed as 
\begin{align}
\frac{d\sigma}{dt} = \frac{1}{64\pi (p_{\mathrm{cm}})^2 s}
\frac{1}{2}\sum_{s_i,s_f}|T|^2,
\label{eq:Def:dsdt}
\end{align}
where $p_{\mathrm{cm}}$ denotes the pion momentum in the
center-of-mass (CM) frame and the nucleon and $\Lambda$ spins
correspond to the $s_i$ and $s_f$, respectively.
Since the Regge amplitudes have a unique virtue that they can
describe consistently the diffractive pattern both at forward and
backward angles as well as the asymptotic behavior with the
unitarity preserved, $d\sigma/dt$ in Eq.~(\ref{eq:Def:dsdt}) should 
comply with the following asymptotic behavior: 
\begin{align}
\frac{d\sigma}{dt}(s \to \infty, t \to 0) \propto s^{2\alpha(t)-2}.
\label{eq:Asym:dsdt}
\end{align}
The asymptotic behavior of $\mathcal{M}_{K^*} (s,t)$ is derived as 
\begin{align}
\lim_{s \to \infty} \sum_{s_i,s_f}
|\mathcal M_{K^*}(s,t)|^2 \propto s^2,
\label{eq:Asym:Ampl1}
\end{align}
which produces a correct asymptotic behavior of the Regge
amplitude. Note that Eq.(\ref{eq:Asym:Ampl1}) is independent of $t$,
which indicates that the amplitude does not change when $t\to 0$.    

It is of great interest to compare the asymptotic behavior of 
the present Regge amplitude with those of other reactions such as
$\gamma N \to K \Lambda$~\cite{Guidal:1997hy} and  $\pi N \to K^*
\Lambda$~\cite{Kim:2015ita}.  In these cases, the amplitudes with
vector-meson exchange include the antisymmetric tensor
that greatly reduces the amplitudes at very forward angles. Explicitly, 
$\mathcal{M}_{K^*}$ for these two reactions behaves as
\begin{align}
\lim_{s \to \infty} \sum_{s_i,s_f}
|\mathcal M_{K^*}(\gamma N \to K \Lambda,\pi N \to K^* \Lambda)|^2
  \propto  s^2 t,
\label{eq:Asym:Ampl}
\end{align}
which is very much distinguished from the present 
reaction, which has the behavior of Eq.~(\ref{eq:Asym:Ampl1}). 

To obtain the $K^*$-meson trajectory in Eq.~(\ref{eq:RegAmpl1}), we 
follow Ref.~\cite{Brisudova:2000ut} in which the so-called 
``square-root'' trajectory is used
\begin{align}
\alpha(t) = \alpha(0) + \gamma [ \sqrt{T} - \sqrt{T-t} ],
\label{eq:SRTraj}
\end{align}
where $\gamma$ is the universal slope and $T$ the scale parameter being
different for each trajectory.
Equation~(\ref{eq:SRTraj}) can be approximated to a linear form
\begin{align}
\alpha(t) = \alpha(0) + \alpha't,
\label{eq:LinTraj}
\end{align}
in the limit $-t \ll T$ with the slope $\alpha'=\gamma/2\sqrt{T}$.
In Ref.~\cite{Brisudova:2000ut}, the value of $\gamma$ and $\sqrt{T}$
were determined in the case of the $\rho$ Reggeon as follows:
\begin{align}
\gamma = 3.65 \pm 0.05\,\mathrm{GeV^{-1}},\,\,\,
\sqrt{T_\rho}=2.46 \pm 0.03 \,\mathrm{GeV}.
\label{UnivPara}
\end{align}
Following this method, we are able to find the corresponding
value of $\sqrt{T}$ for the $K^*$-meson trajectory. The additivities
of intercepts and of inverse slopes are given
as~\cite{Kaidalov:1980bq,Boreskov:1983bu}
\begin{align}
2\alpha_{\bar s q}(0) &= \alpha_{\bar{q}q}(0) +
                       \alpha_{\bar{s}s}(0),\cr  
2/\alpha'_{\bar s q} &= 1/\alpha'_{\bar q q} + 1/\alpha'_{\bar s s},
\label{eq:TrajRela1}
\end{align}
where the $\alpha_{\bar q q}(t)$, $\alpha_{\bar s q}(t)$, and 
$\alpha_{\bar s s}(t)$ are the trajectories corresponding to $\rho$,
$K^*$, and $\phi$ mesons, respectively. Thus, using
Eq.~(\ref{eq:TrajRela1}), we can find the values of 
$\alpha(0)$, $\sqrt{T}$ and $\alpha'$ for the $\phi$ trajectory. 
We summarize the parameters obtained for the vector-meson trajectories
in Table~\ref{tab:1}~\cite{Brisudova:2000ut,Titov:2008yf}.
\begin{table}[h]
\begin{tabular}{c|ccc}
\hline\hline
&$\alpha(0)$&$\sqrt{T}$\,[GeV]&$\alpha'\,[\mathrm{GeV}^{-2}]$ 
\\\hline
$\bar{q}q(\rho)$&0.55&2.46&0.742 \\
$\bar{s}q(K^*)$&0.414&2.58&0.707 \\
$\bar{s}s(\phi)$&0.27&2.70&0.675 \\
\hline\hline
\end{tabular}
\caption{The vector-meson trajectories in the strangeness
 sector~\cite{Brisudova:2000ut,Titov:2008yf}.} 
\label{tab:1}
\end{table}

Once we know all the parameters for the vector-meson Regge
trajectories, we can easily derive the energy scale parameter
$s_{K^*}^{\pi N : K \Lambda}$ in
Eq.~(\ref{eq:RegAmpl1})~\cite{Kaidalov:1980bq,Boreskov:1983bu}  
\begin{align} 
(s_{K^*}^{\pi N : K \Lambda})^{2(\alpha_{K^*}(0)-1)}
=(s^{\pi N})^{\alpha_\rho(0)-1} \times (s^{K \Lambda})^{\alpha_\phi(0)-1}.       
\label{eq:EneScaPara1}
\end{align}
Using the QGSM~\cite{Kaidalov:1980bq,Boreskov:1983bu}, we find the scale 
parameters $s^{\pi N}$ and $s^{K \Lambda}$: $s^{\pi N} \simeq 1.5\,
\mathrm{GeV^2}$ and $s^{K \Lambda} \simeq 1.76\, \mathrm{GeV^2}$. 
Thus, $s_{K^*}^{\pi N : K \Lambda}$ is obtained as $s_{K^*}^{\pi N : K
  \Lambda} \simeq 1.66\, \mathrm{GeV^2}$ by Eq.~(\ref{eq:EneScaPara1}).  
Note that the $t$-channel energy scale parameter $s_{K^*}^{\pi N : K
  \Lambda}$ is the same as that of $s_{K^*}^{\pi N : K^* \Lambda}$ 
given in the reaction $\pi N \to K^* \Lambda$~\cite{Kim:2015ita}, 
because of the same flavor content, $s^{K\Lambda} = s^{K^*\Lambda}$. 

\subsection{Strangeness production $\pi^- p \to K^0\Lambda$:
$\Sigma$ Reggeon exchange}
We now turn to the nonplanar diagram in Fig.~\ref{FIG1}(b).
Though the vector-meson Reggeon exchange contributes to the amplitude
dominantly, baryon exchange also comes into play in describing the
experimental data at backward angles. The effective Lagrangians for
the $\Sigma$ exchange are given as 
\begin{align}
\mathcal L_{K N \Sigma} &=
\frac{g_{K N \Sigma}}{M_N + M_\Sigma} \bar N  \gamma^\mu \gamma_5 
\bm{\tau}\cdot\bm{\Sigma} \partial_\mu  K + \mathrm{H.c.},           \cr   
\mathcal L_{\pi \Sigma \Lambda} &=
\frac{g_{\pi \Sigma \Lambda}}{M_\Lambda + M_\Sigma} \bar \Lambda \gamma^\mu
\gamma_5 \partial_\mu \bm{\pi}\cdot\bm{\Sigma} + \mathrm{H.c.},
\label{eq:Lag2}
\end{align}
where $\Sigma$ represents the lowest-lying $\Sigma(1190,1/2^+)$
hyperon. The coupling constants $g_{KN\Sigma}$ and
$g_{\pi\Sigma\Lambda}$ are taken from the Nijmegen model
(NSC97a)~\cite{Stoks:1999bz} 
\begin{align}
g_{K N \Sigma} = 4.09, \,\,\, g_{\pi \Sigma \Lambda} = 11.9.
\label{eq:Coupl2}
\end{align}

As done in the $t$-channel Reggeon exchange, we can construct the
$u$-channel Regge amplitude~\cite{Kim:2015ita}
\begin{align}
T_\Sigma (s,u)& = C_\Sigma(u)
\mathcal M_\Sigma(s,u)
\left( \frac{s}{s_\Sigma^{\pi N : K \Lambda}} \right)^{\alpha_\Sigma(u)-\frac{1}{2}} 
\Gamma \left( \frac{1}{2}-\alpha_\Sigma(u) \right) \alpha_\Sigma', 
\label{eq:RegAmpl2}
\end{align}
where $C_{\Sigma}(u)$ is the scale factor in the $u$ channel, defined
as 
\begin{align}
C_\Sigma(u) = \frac{a_u}{(1 - u / \Lambda^2)^2}.
\label{eq:ResFac}
\end{align}
To avoid ambiguity, we use the same value of the cut-off mass
$\Lambda$ as in the $t$ channel. The parameter $a_u$ is fitted to the
experimental data: $a_u=2.00$. The amplitude $\mathcal M_\Sigma$ is
written as 
\begin{align}
\mathcal M_\Sigma(s,u)= g_{\pi \Sigma \Lambda} g_{K N \Sigma} \bar
  u_\Lambda (\rlap{/}{q_u}-M_{\Sigma}) u_N,   
\label{eq:Ampl2}
\end{align}
where $q_u$ denotes the momentum transfer in the $u$ channel,
expressed as $q_u = p_2-k_1$.

The $u$-channel amplitude should obey the following asymptotic
behavior 
\begin{align}
\frac{d\sigma}{du}(s \to \infty, u \to 0) \propto s^{2\alpha(u)-2}.
\label{eq:Asym:dsdu}
\end{align}
The unpolarized sum of the sqrared amplitude in Eq.(\ref{eq:Ampl2}) is
shown to be proportional to $s$, as $s\to\infty$
\begin{align}
\lim_{s \to \infty} \sum_{s_i,s_f}
|\mathcal M_{\Sigma}(s,u)|^2 \propto s,
\label{eq:Asym:Ampl2}
\end{align}
which satisfies the general asymptotic behavior of the Regge
amplitude.

Let us consider the Regge parameters in the $u$ channel.
The $\Sigma$ trajectory we use is given as~\cite{Storrow:1984ct}
\begin{align}
\alpha_\Sigma (u) = -0.79 + 0.87 u.
\label{eq:SigmaTraj}
\end{align}
Since the QGSM is applicable only to the planar 
diagram~\cite{Kaidalov:1980bq,Boreskov:1983bu}, we cannot rely on it to 
determine the energy scale parameter $s_\Sigma^{\pi N : K \Lambda}$ in the $u$ 
channel. Instead, we examine carefully a general phenomenological 
relation between the energy scale parameters ($s_0$) and the reaction
thresholds ($s_{\mathrm{th}}$), based on previous and present 
works in both the strangeness and charm sectors. Since we already know
the values of $s_0$ in the $t$ channel, we attempt to find a relation
between $s_0$ and $s_{\mathrm{th}}$ by defining $s_0 /
\sqrt{s_{\mathrm{th}}} = \beta$. Then, we get $s_{K^*}^{\pi
  N : K \Lambda} / \sqrt{s_{\mathrm{th}}^s}  = 1.0\,\mathrm{GeV}$ and
$s_{D^*}^{\pi N : D \Lambda_c} / \sqrt{s_{\mathrm{th}}^c} =
1.1\,\mathrm{GeV}$
($s_{D^*}^{\pi N : D \Lambda_c}$ will be calculated in the next
subsection), where $\sqrt{s_{\mathrm{th}}^s} = M_K + M_\Lambda 
= 1.61\,\mathrm{GeV}$ and $\sqrt{s_{\mathrm{th}}^c} = M_D +
M_{\Lambda_c} = 4.16\,\mathrm{GeV}$. Interestingly, the value of
$\beta$ is always kept to be around $1\,\mathrm{GeV}$, being independent 
of a type of the reactions. For example, we find also 
$\beta\approx 1\,\mathrm{GeV}$ for the $\bar p p \to \bar
\Lambda \Lambda\,(\bar \Lambda_c^+ \Lambda_c^+)$ and $\bar p p \to
\bar K K\,(\bar D D)$ reactions~\cite{Titov:2008yf}. 
Thus, it is reasonable to choose
$s_0$ in such a way that the value of $\beta$ is kept to be equal to
$1\,\mathrm{GeV}$. The energy scale parameters in the $u$ channel are
taken to be $s_\Sigma^{\pi N : K \Lambda} / \sqrt{s_{\mathrm{th}}^s} =
s_{\Sigma_c}^{\pi N : D \Lambda_c} / \sqrt{s_{\mathrm{th}}^c} =
1.0\,\mathrm{GeV}$ in the present calculation.

\subsection{Charm production $\pi^- p \to D^- \Lambda_c^+$}
We extend the study of the strangeness production to the charm production
$\pi^- p \to D^- \Lambda_c^+$ just by substituting charm hadrons for the
strangeness ones, i.e., $K \to \bar D$, $K^* \to \bar D^*$, $\Lambda \to 
\Lambda_c^+$, and $\Sigma^0 \to \Sigma_c^+$. 
\begin{table}[htp]
\begin{tabular}{c|ccc}
\hline\hline
&$\alpha(0)$&$\sqrt{T}$\,[GeV]&$\alpha'\,[\mathrm{GeV}^{-2}]$ 
\\\hline
$\bar{q}q(\rho)$&0.55&2.46&0.742 \\
$\bar{c}q(D^*)$&-1.02&3.91&0.467 \\
$\bar{c}c(J/\psi)$&-2.60&5.36&0.340 \\
\hline\hline
\end{tabular}
\caption{Vector-meson trajectories in the charm
sector~\cite{Brisudova:2000ut,Titov:2008yf}.}
\label{tab:2}
\end{table}
The $D^*$-meson trajectory in $t$-channel exchange 
can be found as in the case of the strangeness production. The
relevant results are listed in
Table~\ref{tab:2}~\cite{Brisudova:2000ut,Titov:2008yf}.
Equation~(\ref{eq:EneScaPara1}) is also modified as
\begin{align} 
(s_{D^*}^{\pi N : D \Lambda_c})^{2(\alpha_{D^*}(0)-1)}
=(s^{\pi N})^{\alpha_\rho(0)-1} \times (s^{D \Lambda_c})^{\alpha_{J/\psi}(0)-1},
\end{align}
and we get:
$s^{\pi N} \simeq 1.5\, \mathrm{GeV^2},\, 
s^{D \Lambda_c} \simeq 5.46\, \mathrm{GeV^2},\,$ and
$s_{D^*}^{\pi N : D \Lambda_c} \simeq 4.748\, \mathrm{GeV^2}$.
Lastly, the $\Sigma_c$ trajectory is found to be
\begin{align}
\alpha_{\Sigma_c} (u) = -2.23 + 0.532 u,
\label{eq:SigmacTraj}
\end{align}
as done similarly in Ref.~\cite{Titov:2008yf} where the $\Lambda_c$ 
trajectory was calculated to be $\alpha_{\Lambda_c} (u) = -2.09 +
0.557 u$. 

Note that for the scale factors in Eqs.~(\ref{eq:RegAmpl1})
and~(\ref{eq:RegAmpl2}) the same values will be used in this charm 
production to avoid ambiguity. We also consider the same values of the
coupling constants for the corresponding vertices  
such that we can compare the magnitudes of the observables for the charm 
production with those for the strangeness production. 

\section{Results and discussion}
\vspace{1.5em}
\begin{figure}[htp]
\centering
\includegraphics[scale=0.35]{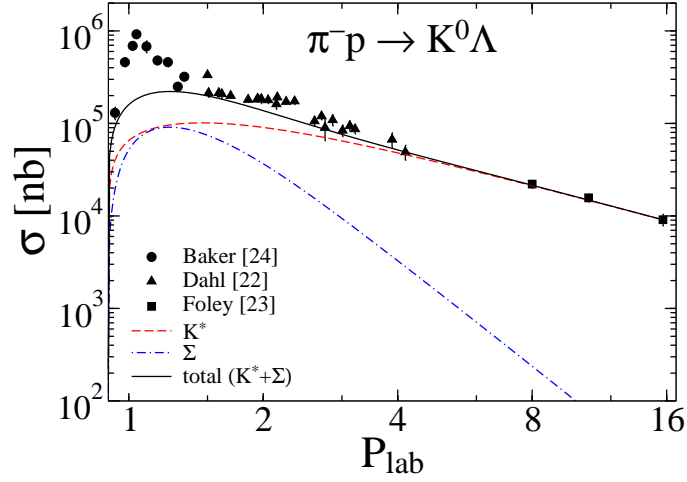}
\caption{(color online) Total cross section with each contribution is
plotted as a function of $s/s_{\mathrm{th}}$ for the 
$\pi^- p \to K^0 \Lambda$. The experimental data are taken
from Ref.~\cite{Baker:1978qm} (circle), Ref.~\cite{Dahl:1969ap}
(triangle), and Ref.~\cite{Foley:1973ve} (square).}
\label{fig:2}
\end{figure}
Figure~\ref{fig:2} shows the numerical results of the total cross section 
for the $\pi^- p \to K^0 \Lambda$ as a function of $s/s_{\mathrm{th}}$,
where $s_{\mathrm{th}}$ designates the threshold value, i.e., 
$s_{\mathrm{th}} = (M_K + M_\Lambda)^2 = 2.60\,\mathrm{GeV^2}$.
They are in good agreement with the experimental data at
intermediate~\cite{Dahl:1969ap} and high~\cite{Foley:1973ve} energies.
On the other hand, the present results seem underestimated near
threshold, compared with the data~\cite{Baker:1978qm}. Note that the
resonance contribution plays a dominant role in explaining the data
near threshold. We do not take into account the nucleon resonances in
the $s$ channel in the present work, since we are mainly interested in
studying the order of magnitude of the charm production in comparison
with that of the strangeness production. 
As drawn in Fig.~\ref{fig:2}, the $K^*$ and $\Sigma$ Reggeons
have comparable effects on the total cross section in the lower energy
region. However, as $P_{\mathrm{lab}}$ increases, the contribution of
the $\Sigma$ falls off much faster than that of the $K^*$. Thus,
the dependence of the total cross section on $P_{\mathrm{lab}}$ is
mainly governed by the $t$-channel process. The reason can be found in
the fact that the intercept of Regge trajectory $\alpha(0)$ for the
$K^*$ Reggeon is much larger than that for the $\Sigma$ Reggeon.

\begin{figure}[htp]
\centering
\includegraphics[scale=0.62]{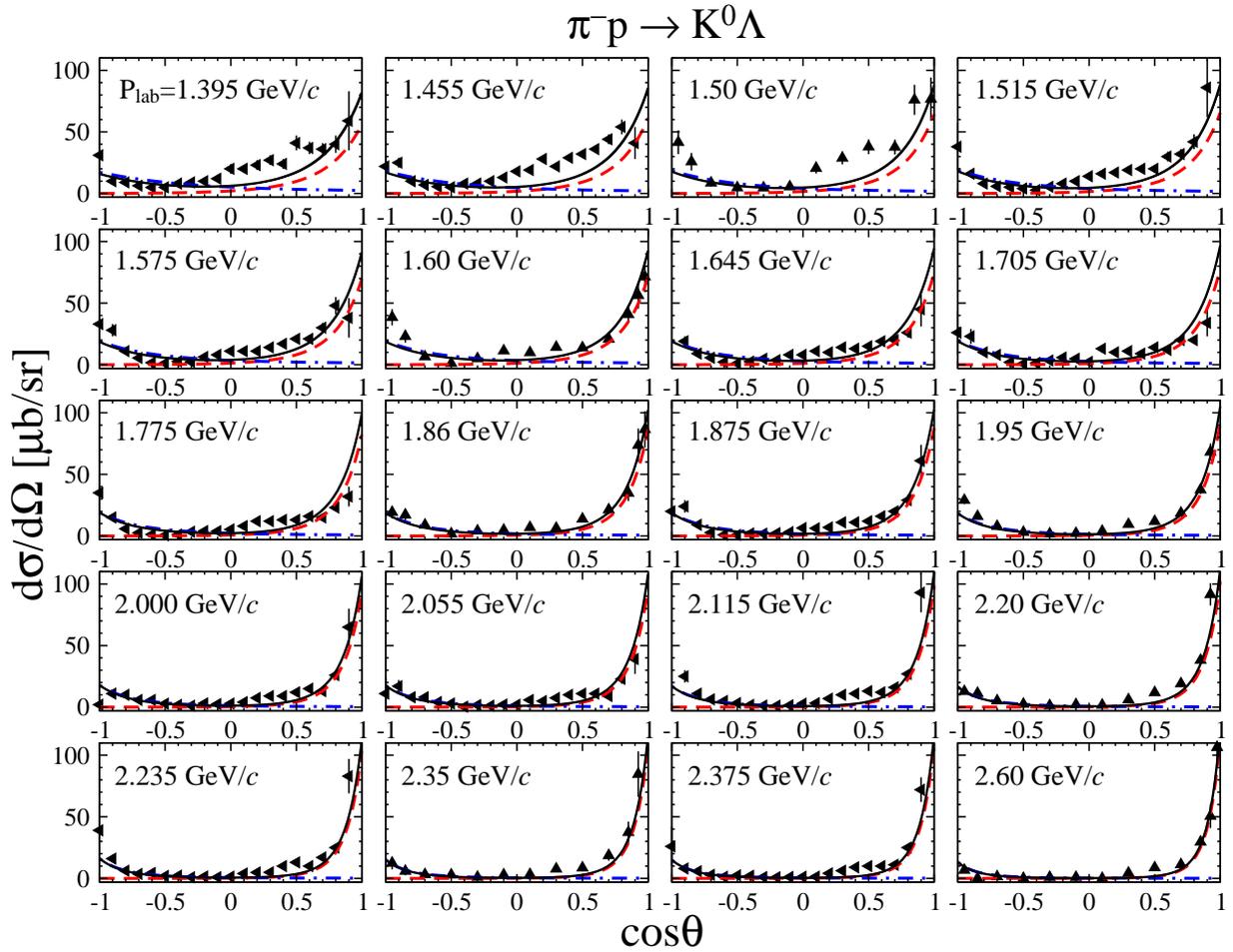}
\caption{(color online) Differential cross sections with each contribution
are plotted as functions of $\cos\theta$ for the 
$\pi^- p \to K^0 \Lambda$ in the range of 
1.395 GeV $\le$ $P_{\mathrm{lab}}$ $\le$ 2.60 GeV.
The experimental data are taken from Ref.~\cite{Dahl:1969ap} 
(triangle up) and Ref.~\cite{Saxon:1979xu} (triangle left).
The notation is the same as Fig.~\ref{fig:2}.}
\label{fig:3}
\end{figure}
\begin{figure}[htp]
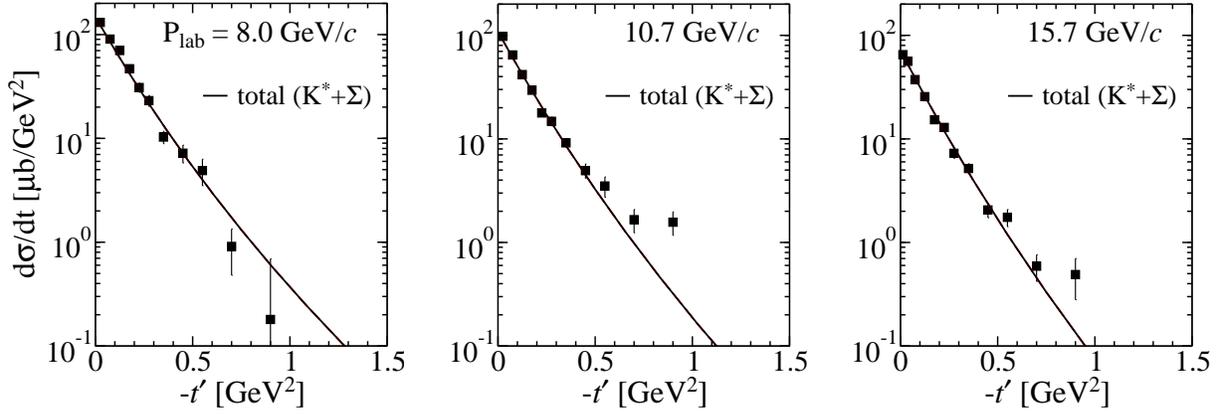

\includegraphics[width=5.30cm]{FIG4a.eps}\hspace{1.1em}
\includegraphics[width=4.80cm]{FIG4b.eps}\hspace{1.1em}
\includegraphics[width=4.80cm]{FIG4c.eps}
\caption{Differential cross sections for the $\pi^- p \to K^0 \Lambda$ 
are plotted as functions of $-t'$ at three different pion momenta 
($P_{\mathrm{lab}}$).
The experimental data are taken from Ref.~\cite{Foley:1973ve}.}
\label{fig:4}
\end{figure}

In Fig.~\ref{fig:3}, we depict the results of the differential cross
sections $d\sigma/d\Omega$ as functions of $\cos\theta$, given 20
different values of $P_{\mathrm{lab}}$. They are in good agreement
with the experimental data~\cite{Dahl:1969ap, Saxon:1979xu}, when
$P_{\mathrm{lab}}$ is larger than 1.6 GeV. The discrepancy of 
our results from the data below 1.6 GeV arises from the same
reason that we have not included the nucleon resonances in the $s$
channel. Figure~\ref{fig:4} displays the results of the
differential cross section $d\sigma/dt$ as functions of $-t'$ 
defined as $t'=t-t_{\mathrm{min}}$, where $-t_{\mathrm{min}}$
represents the smallest kinematical value of $-t$ at fixed
$P_{\mathrm{lab}}$. The results agree with
the experimental data very well up to $-t'=0.5\,\mathrm{GeV}^2$. We want
to mention that $K^*$ Reggeon exchange dictates the dependence of
$d\sigma/dt$ on $t'$. 

\begin{figure}[htp]
\includegraphics[scale=0.35]{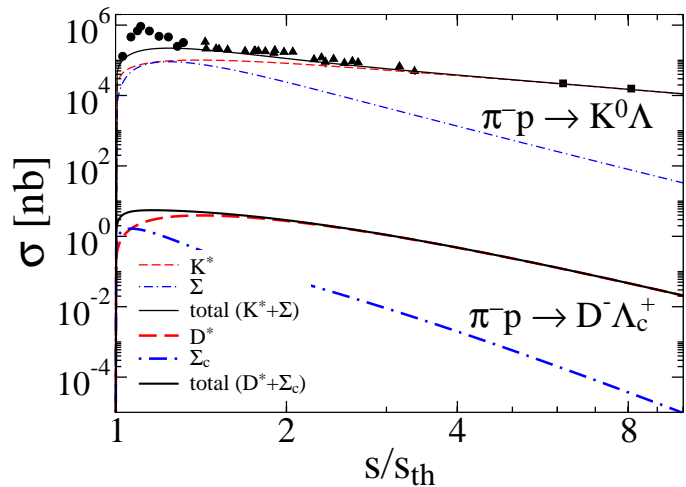}
\caption{(color online) Total cross section with each 
contribution is plotted as a function of $s/s_{\mathrm{th}}$ for the 
$\pi^- p \to D^- \Lambda_c^+$ in comparison with that for the $\pi^-
p \to K^0 \Lambda$. The experimental data are taken from
Refs.~\cite{Baker:1978qm,Dahl:1969ap,Foley:1973ve}.}
\label{fig:5}
\end{figure}
In Fig.~\ref{fig:5}, we draw the results of the total 
cross section for the $\pi^- p \to D^- \Lambda_c^+$ as a function of 
$s/s_{\mathrm{th}}$ in comparison with that for the $\pi^- p \to K^0 \Lambda$. 
Here the threshold value for the charm production is given as
$s_{\mathrm{th}} = (M_D + M_{\Lambda_c})^2 = 17.3\,\mathrm{GeV^2}$.
Each contribution has a similar tendency as in the case of the $\pi^-
p \to K^0 \Lambda$. As $s/s_{\mathrm{th}}$ increases, the total cross
section is almost controlled by the $D^*$ Reggeon exchange. However,
the magnitude of the total cross sections for the charm production is
about $10^4-10^6$ times smaller than that for the strangeness
production. We find the similar results in the study of the
$K^*\Lambda$ and $D^* \Lambda_c^+$ production
reactions~\cite{Kim:2015ita}. 
Both the inptercept $\alpha(0)$ and energy scale parameter $s_0$ are
crucial to determine the total cross section for a corresponding 
process.
While $\alpha_{K^*}(0)$ is given as $\alpha_{K^*}(0)=0.414$ as listed in 
Table~\ref{tab:1}, $\alpha_{D^*}(0)$ is found to be $-1.02$ in 
Table~\ref{tab:2}. 
This difference makes the total cross section fall off faster
than that of the strange production as $s/s_{\mathrm{th}}$ increases.
The suppression of the present result is also seen in
Ref.~\cite{Barger:1975fua} where a simplified Regge model is
employed. It is found that the cross sections are sensitive depending
on the values of the intercept and hadron mass.
The most optimistic estimation suggested 0.5 nb
at peak position with $\alpha_{D^*}(0) = -0.6$~\cite{Barger:1975fua},
whereas the present work yields 4 nb. 

\begin{figure}[htp]
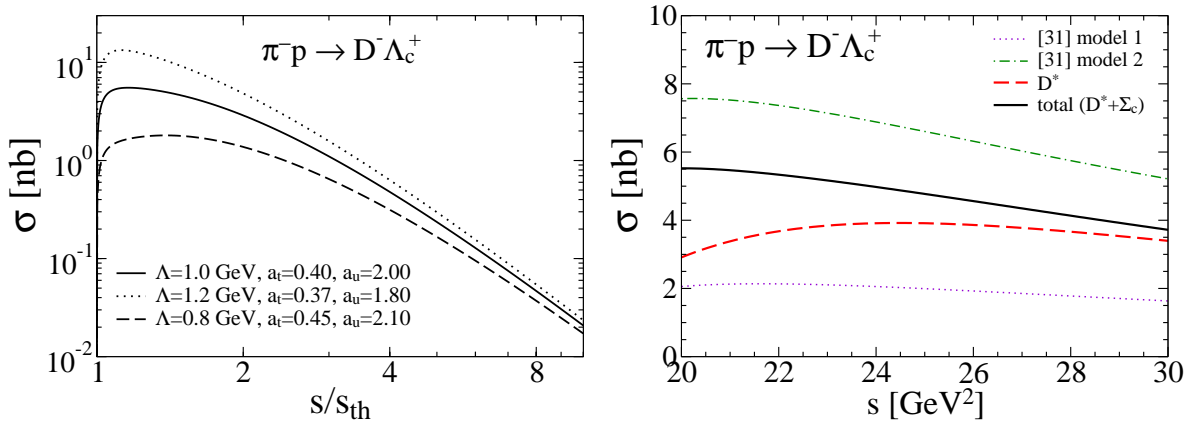

\centering
\includegraphics[scale=0.30]{FIG6a.eps}\,\,\,
\includegraphics[scale=0.30]{FIG6b.eps}
\caption{(color online) Left panel: Total cross section for the 
$\pi^- p \to D^- \Lambda_c^+$ is plotted as a function of $s/s_{\mathrm{th}}$
with different parameter sets in the scale factors. 
Right panel: The present results are compared with those of 
Ref.~\cite{Kofler:2014yka} for the total cross section.
The range of $x$-axis $20\,\mathrm{GeV^2} \leq s \leq 30\,\mathrm{GeV^2}$ 
corresponds to $1.2\,\leq s/s_{\mathrm{th}} \leq 1.7$ in Fig.~\ref{fig:5}.}
\label{fig:6}
\end{figure}
Now we would like to check the uncertainties of the present results 
by using different sets of parameters. 
As mentioned already, we have employed the set 
$(\Lambda = 1\,\mathrm{GeV}, a_t = 0.40, a_u = 2.00)$ for the charm 
production, which is determined from the strangeness sector. 
However, this is not a unique choice.
Within a reasonable range of the cut off $\Lambda$ around 1 GeV, the 
parameter sets 
$(\Lambda = 1.2\,\mathrm{GeV}, a_t = 0.37, a_u = 1.80)$ and 
$(\Lambda = 0.8\,\mathrm{GeV}, a_t = 0.45, a_u = 2.10)$ can equally 
reproduce the strangeness production. 
When we apply these values to the charm production, the total cross 
sections lie in the range $1\, \mathrm{nb}- 13\,\mathrm{nb}$  near 
threshold as shown in the left panel of Fig.~\ref{fig:6}. 
As the production energy increases, the difference becomes smaller 
gradually. It is worthwhile to compare our results
with those from the other work~\cite{Kofler:2014yka} in which the same
reaction $\pi^- p \to D^- \Lambda_c^+$ was investigated within a
generalized parton picture, focusing on the reaction threshold and
forward angle region. As shown in the right panel of
Fig.~\ref{fig:6}, the results of this work lie between those of Model
I and Model II given in Ref.~\cite{Kofler:2014yka}. The $s$ dependence
of the present results look very similar to those of both Model I and
Model II near threshold
($20\,\mathrm{GeV}^2 \le s \le 30\,\mathrm{GeV}^2$). However, 
as $s$ increases, the results of Ref.~\cite{Kofler:2014yka} fall off
rather slowly~\cite{Kofler:2016pc}, which deviates from the present ones. 
It indicates that the models developed in Ref.~\cite{Kofler:2014yka} do 
not satisfy the asymptotic behavior of the cross sections.

\begin{figure}[htp]
\centering
\includegraphics[scale=0.60]{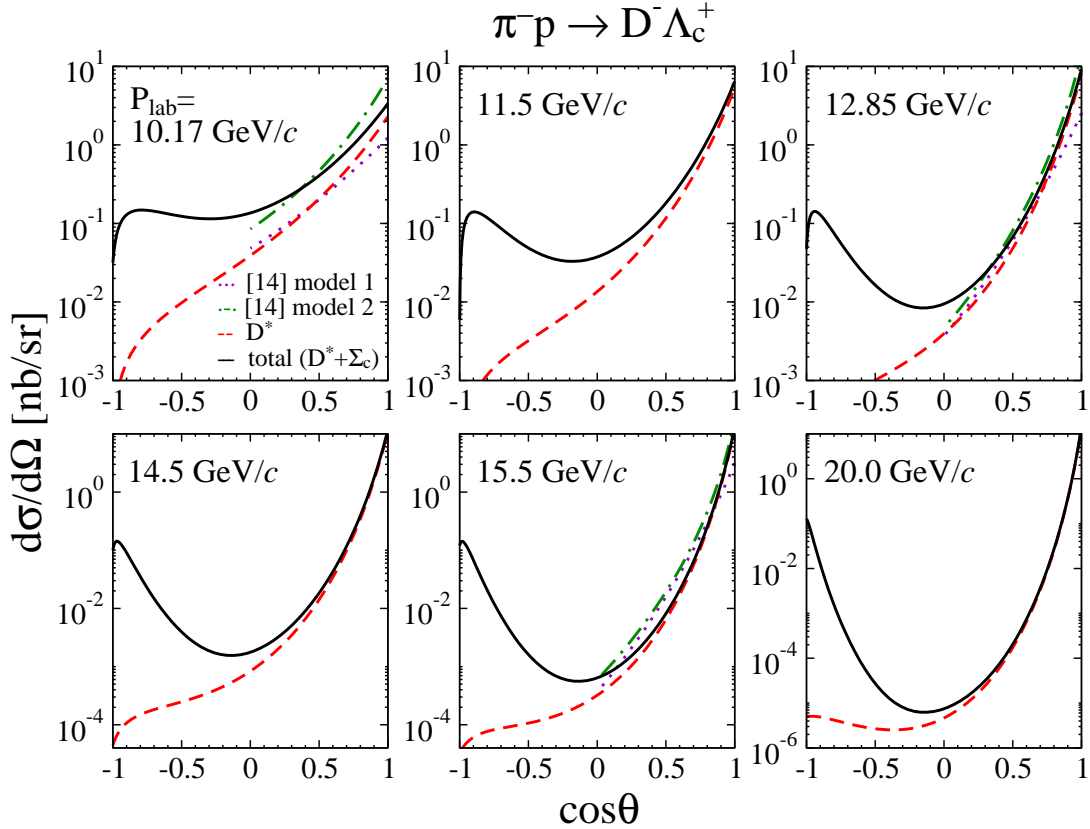}
\caption{(color online) Differential cross sections with each contribution
are plotted as functions of $\cos\theta$ for the 
$\pi^- p \to D^- \Lambda_c^+$ at six different pion momenta 
($P_{\mathrm{lab}}$). The present results are compared with those of
Ref.~\cite{Kofler:2014yka}.} 
\label{fig:7}
\end{figure}
In Fig.~\ref{fig:7}, we continue to compare the present results of the
differential cross sections $d\sigma/d\Omega$ with those of
Ref.~\cite{Kofler:2014yka}. It is interesting to see that at forward
angles both results are in good agreement with each other. However,
the differential cross sections in backward angles were not considered
in the models of Ref.~\cite{Kofler:2014yka}. 
\begin{figure}[htp]
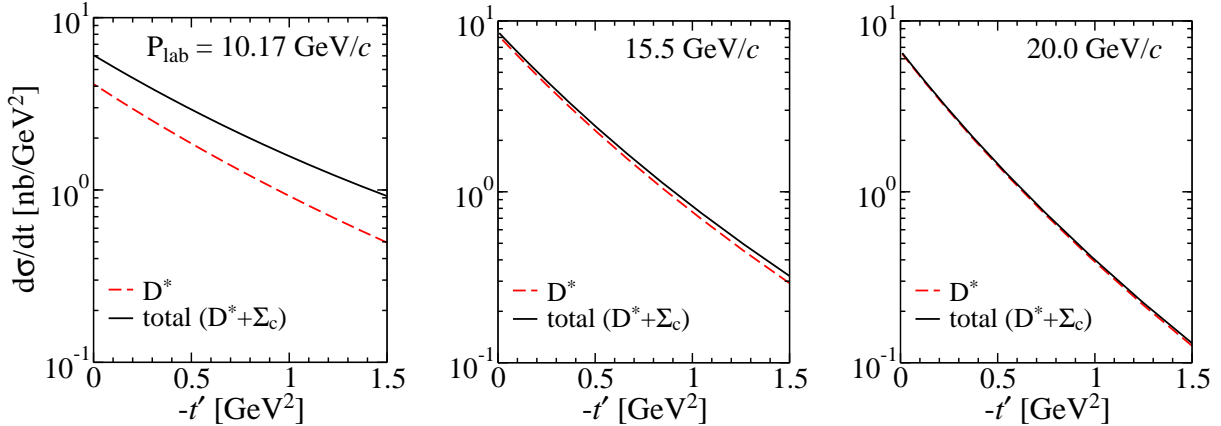

\includegraphics[width=5.30cm]{FIG8a.eps}\hspace{1.1em}
\includegraphics[width=4.80cm]{FIG8b.eps}\hspace{1.1em}
\includegraphics[width=4.80cm]{FIG8c.eps}
\caption{Differential cross sections for the 
$\pi^- p \to D^- \Lambda_c^+$ are plotted as functions of $-t'$ at three 
different pion momenta ($P_{\mathrm{lab}}$).}
\label{fig:8}
\end{figure}
Figure~\ref{fig:8} illustrates the results of the differential cross
sections $d\sigma/dt$ as functions of $-t'$ at three different values
of $P_{\mathrm{lab}}$. The $D^*$ Reggeon governs $t'$ dependence at
higher $P_{\mathrm{lab}}$, whereas the $u$ channel has certain effects
on the $d\sigma/dt$ near threshold. 
\section{Conclusion and Summary}
In the present work, we aimed at investigating the production
mechanism of the $\pi^- p\to K^0 \Lambda$ and $\pi^-p\to
D^-\Lambda_c^+$ reactions, based on a hybridized Regge model. We
replaced the Feynman propagator by the Regge one from the invariant
amplitudes. The Regge amplitudes explain the asymptotic behavior
of the cross sections for the present reactions such that unitarity is
well preserved, whereas the amplitudes from the effective Lagrangians
correctly reproduce the magnitude of the cross sections near
threshold. Combining each virtue of these two different approaches, we
were able to study both the $K\Lambda$ and $D\Lambda_c$ productions
consistently. Having determined the Regge parameters of $K^*$ and
$D^*$ Reggeons by using the quark-gluon string model, and having fixed
those of $\Sigma$ and $\Sigma_c$ Reggeons phenomenologically, we
have computed the total cross sections, the differential cross
sections $d\sigma/d\Omega$ and $d\sigma/dt$. As in the case of the
$K^* \Lambda$ and $D^* \Lambda_c$ productions, we 
have found that the charm production is
 suppressed almost by similar order, i.e., the total
cross sections for the charm production are $10^{4}-10^{6}$ times
smaller than those for the strangeness production
depending on kinematical regions. 
We have compared the present results with those from the other work
near the threshold region and at forward angles. Both the results are  
qualitatively in agreement with each other. 

Anticipating that the J-PARC, the BES-III, the JLAB,
and the LHC facilities will produce a great deal of new experimental
data related to charm physics, we find that it is of paramount
importance to investigate the production mechanism of charmed
hadrons with various probes. Relevant theoretical studies are under
way.  

\section*{Acknowledgments}
We are grateful to S.Kofler for sending us their results. 
This work was supported by the National Research Foundation of
Korea (NRF) grant funded by the Korea government (MSIP)
(No.NRF-2015R1A2A2A04007048). S.H.K acknowledges the support of the
Young Scientist Training Program at the Asia Pacific Center for
Theoretical Physics from the Korea Ministry of Education, Science and
Technology, Gyeongsangbuk-Do and Pohang City.  AH acknowledges the
support in part by Grant-in-Aid for Science Research (C) JP26400273  by
the MEXT. 



\end{document}